\documentclass{nime04}

\begin{document}

\title{A Novel Face-tracking Mouth Controller and its\\
Application to Interacting with Bioacoustic Models}

\numberofauthors{3}

\author{
\alignauthor Gamhewage C. de Silva \\
       \affaddr{ATR MIS Labs}\\
       \affaddr{2-2-2 Hikaridai, Keihanna Science City}\\
       \affaddr{Kyoto, Japan, 619-0288}\\
       \email{chamds@atr.jp}
\alignauthor Tamara Smyth \\
       \affaddr{CCRMA}\\
       \affaddr{Stanford University}\\
       \affaddr{Stanford, CA  94305}\\
       \email{tamara@ccrma.stanford.edu}
\alignauthor Michael J. Lyons\\
     \affaddr{ATR IRC and MIS Labs}\\
       \affaddr{2-2-2 Hikaridai, Keihanna Science City}\\
       \affaddr{Kyoto, Japan, 619-0288}\\
       \email{mlyons@atr.jp}
  }
\date{}
\maketitle
\begin{abstract}
We describe a simple, computationally light, real-time system for tracking the lower face and 
extracting information about the shape of the open mouth from a video sequence. The system
allows unencumbered control of audio synthesis modules by action of the mouth. We report
work in progress to use the mouth controller to interact with a  physical model of sound 
production by the avian syrinx.
\end{abstract}


\section{Introduction}
Several studies have explored the use of the mouth or vocal
tract for controlling audio synthesis \cite{Orio97, Lyons01, Vogt02, Lyons03}.
The motivations for this line of research relate to the role of the mouth in speech, singing,
and facial expression, as previously discussed in some detail \cite{Lyons03}.
The current study builds on previous work by our group on the mouthesizer system 
\cite{Lyons01, Lyons03, Chan03}.  Early versions of the mouthesizer included a 
vision-based head tracking system as a front end, however the head-tracking system was abandoned
in favour of a miniature camera, which greatly simplified the video analysis. In this paper we 
explore a face-tracking system which is simple and computationally light, but quite robust. The main
trade-off is that initialization, though easy, is not fully automatic. The face tracking system we describe
here is different from the vision-based head tracking systems used in other works with musical controllers \cite{Ng02, Merrill03}.
The present work also differs from these latter two studies in that, as with the mouthesizer, we focus
on the action of the mouth.

We are using the mouth controller to interact with a physical model of the avian syrinx. 
Our  interest in bioacoustic models was stimulated by a presentation on the tymbalimba controller 
\cite{Smyth02}. With the tymbalimba, a mechanical
model of the buckling mechanism used by the cicada to generate vibrations is 
employed as an interface to a physical model of sound production by the cicada. 
The  tymbalimba study seemed, to one of the co-authors of the present report (ML), 
to harmoniously blend art, science, and technology to create what might be called an example of a 
{\em poetic technology} \cite{Heidegger}, a technology aimed at revealing nature, rather than
exploiting nature.
In a similar way, the current work is aimed at enabling humans to use a part of their 
vocal apparatus to physically experience a simulation of bird vocalization.
\begin{figure}[t]
\begin{center}\includegraphics[width=3in]{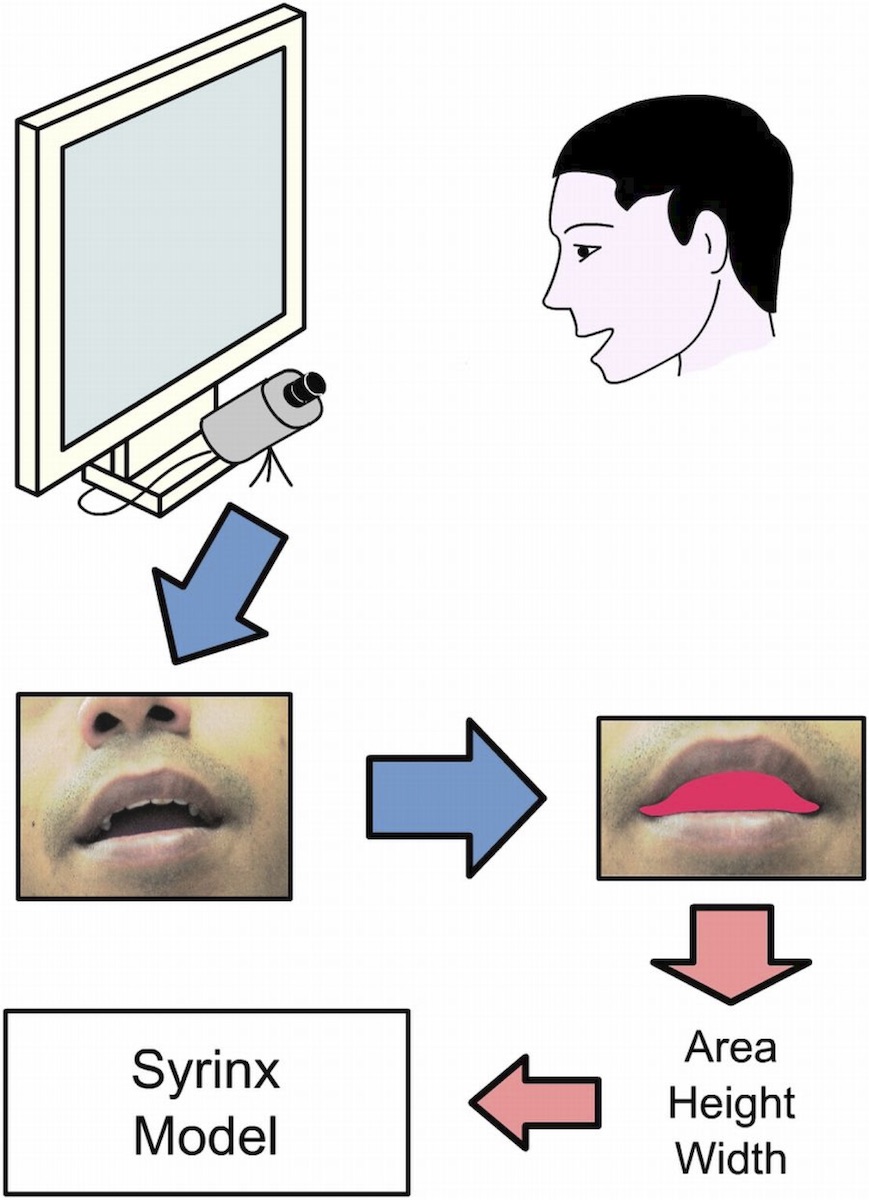}\end{center}
\caption{Overview of the mouth controller.}
\label{fig1}
\end{figure}

\section{Mouth Controller}
Figure 1 gives an overview of the head-tracking mouth controller.  A camera positioned 
below the computer monitor is used to acquire the images of the lower 
region of the user's face. Tracking is iniated manually by roughly positioning
the face so that the nostrils lie in the top central portion of the image while
clicking the left mouse button.
 The nostril detection subsystem detects the centers of the nostrils and 
estimates parameters related to their position and orientation. These 
parameters are used to track nostrils in the subsequent image frames and 
determine the region of the image containing the mouth. The 
mouth cavity is segmented and shape features of the shadow area of the mouth cavity are 
passed to the bioacoustic physical model. The coordinates of the point lying between
the nostril centers is also sent, however we are not currently using these parameters
to control sound synthesis.

A robust and computationally simple approach to nostril detection
and tracking is used. The method is a modified version of the algorithm
proposed by Petajan \cite{Petajan96}, and uses gradients of pixel
intensities as features. The nostrils are cavities, not surfaces, and under
a wide range of lighting conditions they appear dark relative to the surrounding area
of the face.  For an upright face, a small window of the image containing the
nostrils exhibits characteristic patterns of intensity variation in
the horizontal and vertical directions. These patterns are prominently
seen in horizontal and vertical image projections (Figure 2) and can be used
to infer the approximate  location of the nostril centers.

Nostril detection is initialized by positioning the nose
in a specified rectangular region of the image (Figure 2(a)) and clicking the
left mouse button. In practice this requires little effort but still significantly
reduces the complexity of the nostril detection step.
\begin{figure}[t]
\begin{center}\includegraphics[width=3in]{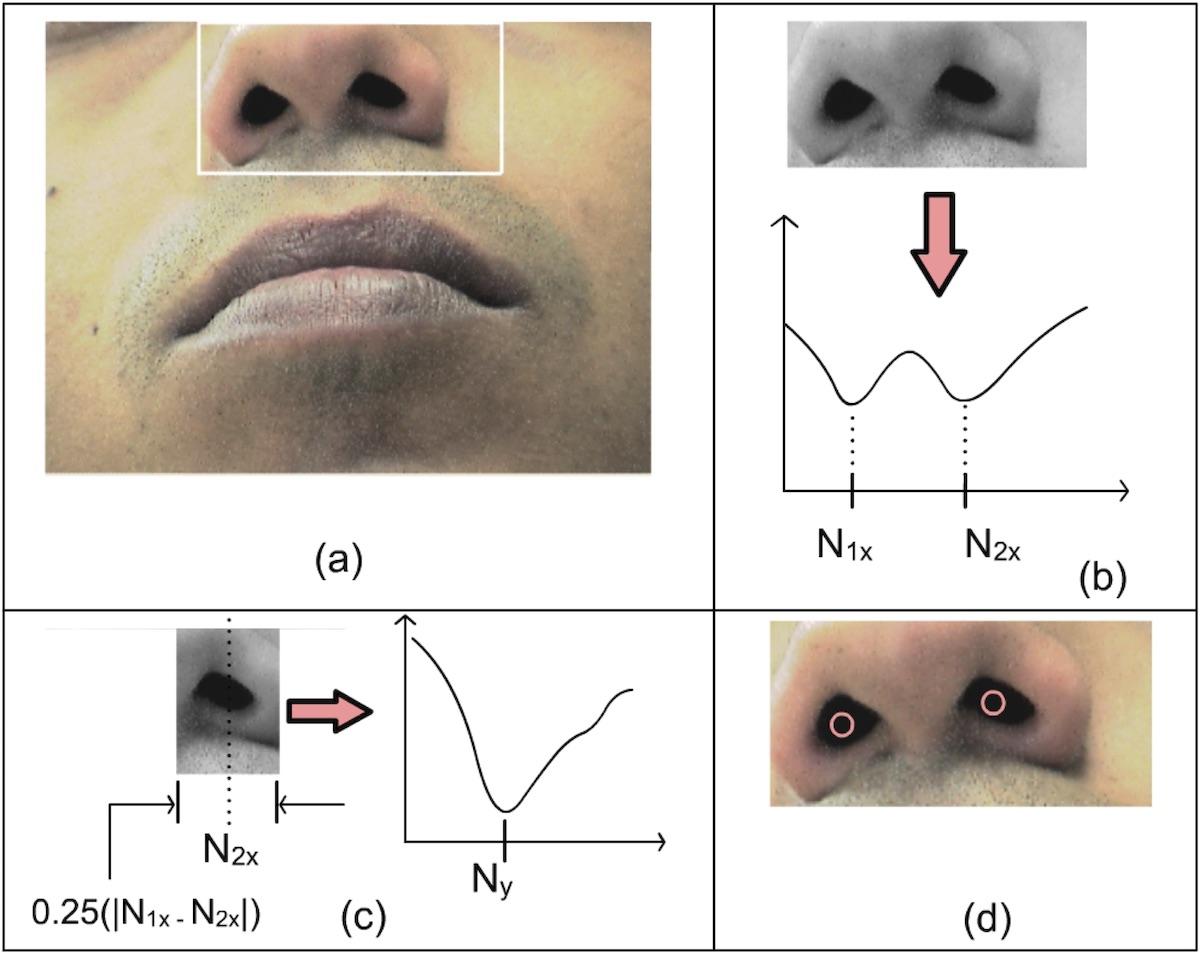}\end{center}
\caption{Steps involved in nostril detection.}
\end{figure}
The subimage containing the nostils is projected onto horizontal and vertical
directions and the projection smoothed by  low pass filtering.
A characteristic pattern with two local minima, corresponding approximately to the two nostrils,
can be observed in the projection. 
The horizontal and vertical coordinates of the centers of the nostrils,  
$N_{1x}$, $N_{2x}$, $N_{1y}$, and $N_{2y}$,
can be estimated from the projections (Figure 2(b) - (d)).

The coordinates of the determined nostril centers $N_1 = (N_1x, N_1y)$ and 
$N_2 = (N_2x, N_2y)$ can be used to determine the distance between
the nostril centers $D_N$, the mid-point between the nostril centers,
$C_N$, and the angle between the line joining the nostril centers and the horizontal axis, $A_N$.
These three parameters determine the window used for nostril detection in the next frame (Figure 3,
top rectangle).

Nostril tracking proceeds in a similar fashion to nostril detection, with the following changes but
the nostril search window is rotated by angle $-A_N$ about $C_N$ before the samples are extracted. This 
improves the accuracy of the measured parameters. $D_N$ and $A_N$ are smoothed by using a weighted sum of the 
previous value and current value to temporally smooth the motion of the search window. Also, the position of 
$C_N$ is predicted  by assuming contant velocity over three consecutive frames, according to he following equation:
$$ C_{N(t+1)} = C_{N(t)} + \alpha \{ C_{N(t)} - C_{N(t-1)}\}. $$  
The nostril locations are also used to determine a region containing the
mouth (Figure 3, lower rectangle). The window is rotated before the subimage
is extrated for processing. This simplifies the task of calculating the orientation
dependent height and width of the mouth. Some portions of the mouth region can be 
outside of the image boundary, but this does not affect mouth cavity segmentation.
\begin{figure}[bt]
\begin{center}\includegraphics[width=3in]{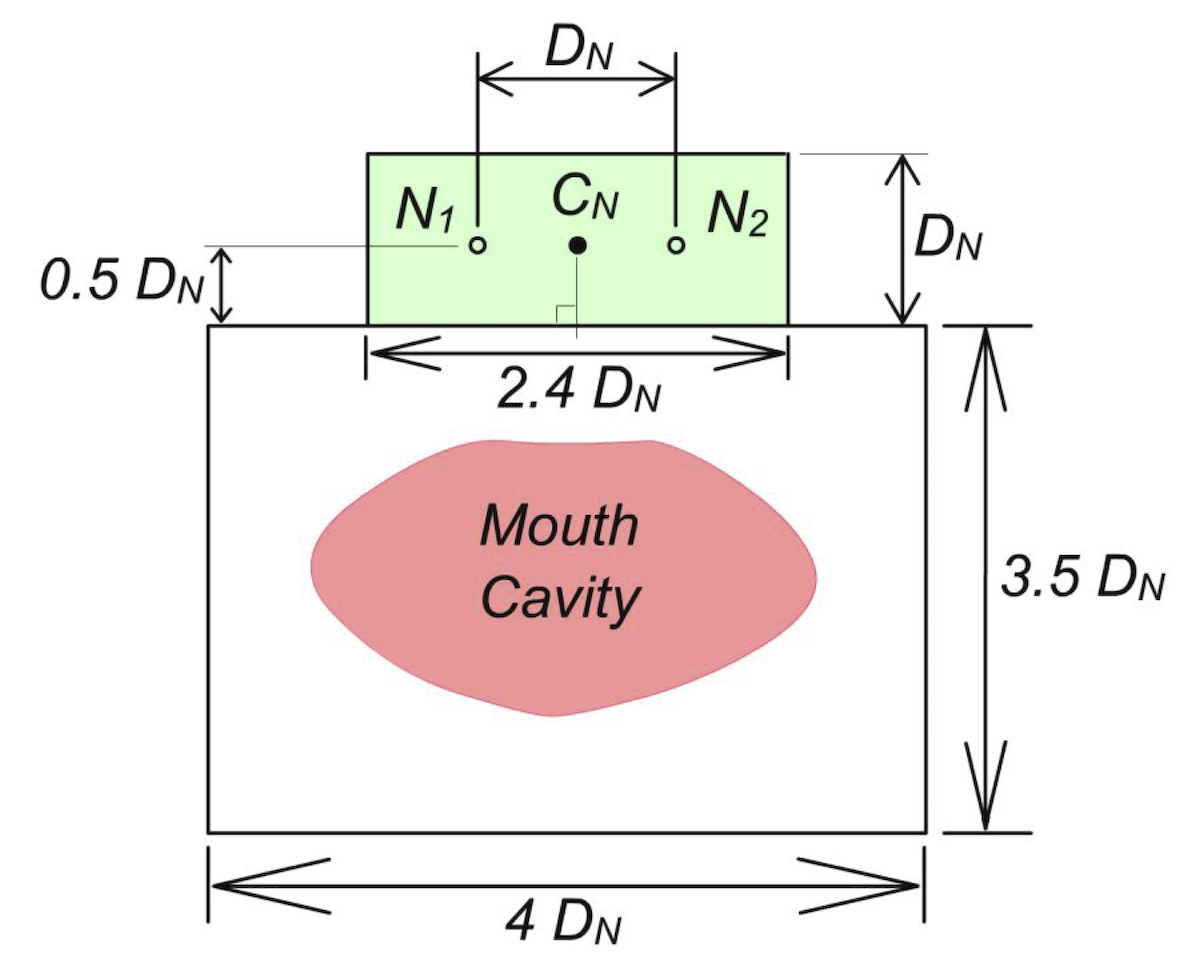}\end{center}
\caption{Upper rectangle: nostril search region. Lower rectangle: mouth region.}
\label{fig4}
\end{figure}

Colour and intensity thresholding are used to segment the  shadow
area inside the open mouth, as with  the  headworn mouthesizer \cite{Lyons01, Lyons03, Chan03}.
The mouth appears as a dark, relatively red region in the image. Pixels
with red component above a certain threshold and intensity less than
another threshold are selected as beloning to the shadow in the mouth cavity.
Each threshold can be controlled with a slider by the user, while observing a visualization
of the segmented mouth region on the screen, to acheive a continuous region.

A  voting algorithm is used to reduce noise. Each segmented pixel is cleared if there are less 
than 4 segmented pixels in a rectangular neighborhood of width 5 and height 
3. Pixels in the mouth region, which are not  segmented by thresholding, 
are added to the segmented region if there are more than 4 
segmented pixels in a rectangular neighborhood of the same size. After 
smoothing, the largest connected blob is selected as corroesponding to the mouth 
cavity region.

Several shape features of the chosen blob are calculated. The total
number of the pixels in the blob is proportional to the {\bf area}, $A$ of the shadow
region of the mouth cavity. The standard deviation of the vertical coordinate
of pixels in the blob is proportional to the {\bf height},$H$ of the mouth cavity.
The standard deviation of the horizontal coordinate of pixels in the blob is
proportional to the {\bf width}, $W$, of the mouth cavity. The width and height of the
mouth cavity could also be estimated using the bounding box of the blob.
However, the standard deviations, which is a function of all pixel coordinates
is less sensitive to noise. The {\bf aspect ratio}, of the mouth cavity region is 
given by $R = H/W$.

The parameters can change when the user moves their head, even if the mouth 
shape does not change. Cancellation of this effect would require information about
the three dimensional shape of the face as well as pose and range information. However
since the system is used interactively, the user can control the pose and range of their
face according to auditory feedback from the system. In practice, we envision using
this system in one of three possible ways: with a desktop camera; with a camera on a stand
such as a microphone stand \cite{Hewitt03}; with a handheld camera. Presently, the area, height, width, and 
aspect ratio, are ouput as MIDI control changes. Other methods of interfacing with synthesis
modules are being investigated.

\section{Model of the Avian Syrinx}
In this work, we focus on the application of the system to control of an example of a bioacoustic
physical model of the avian syrinx developed by Smyth and Smith. 
This section summarizes the model. For further details we refer the reader to
the cited works on the development of the syrinx model.

The bird's airway consists of a trachea which divides into the left
and right bronchus at its base.  Within each bronchus there is a
membrane forming a pressure-controlled valve just before its junction
with the trachea.  During voiced song, the membrane is set into motion
by air flow, vibrating at a frequency determined partly by the mass
and tension of the membrane and partly by the resonance of the upper
bronchus and trachea to which it is connected \cite{flet99}.  The
neural control of the muscles surrounding the syrinx, the pressure in
the interclavicular air sac which encases the syrinx, and the bird's
respiratory mechanics all greatly contribute to how sound is modulated
by the syrinx \cite{brackenbury}.
\begin{figure}[b]
\centering
\epsfig{file=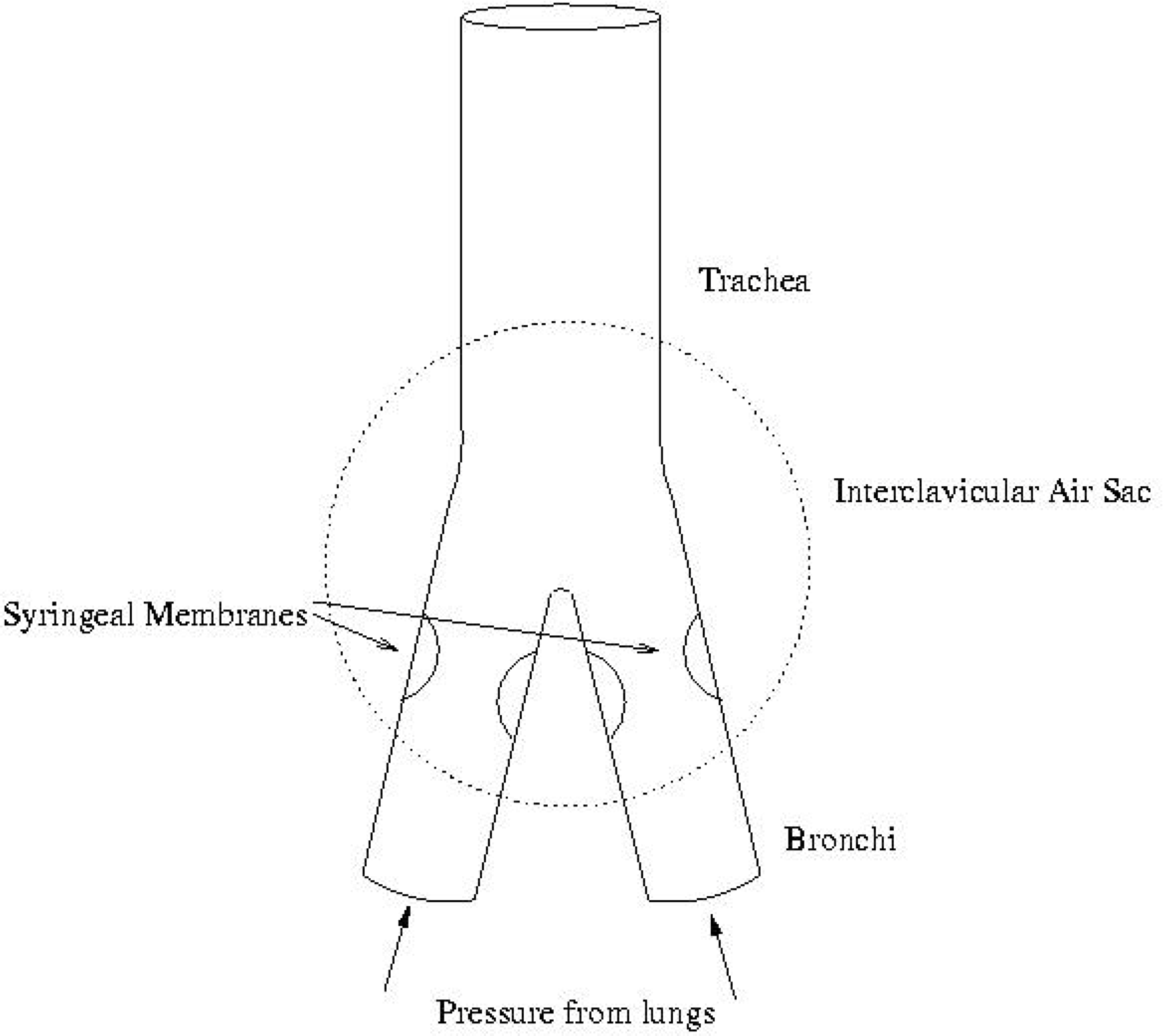, width=2.5in}
\caption{A simplified diagram of a syrinx.}
\label{syrinx}
\end{figure}
When the membrane is set into motion, variable heights are formed
within the valve channel creating a constricted aperture through which
air must flow.  The model of the valve displacement and the resulting
pressure through the constriction is developed following an acoustic
model by Fletcher \cite{flet88} and is based on the mechanical
properties of the membrane and the Bernoulli equation for air flow.
  
\begin{figure}[t]
\centering
\epsfig{file=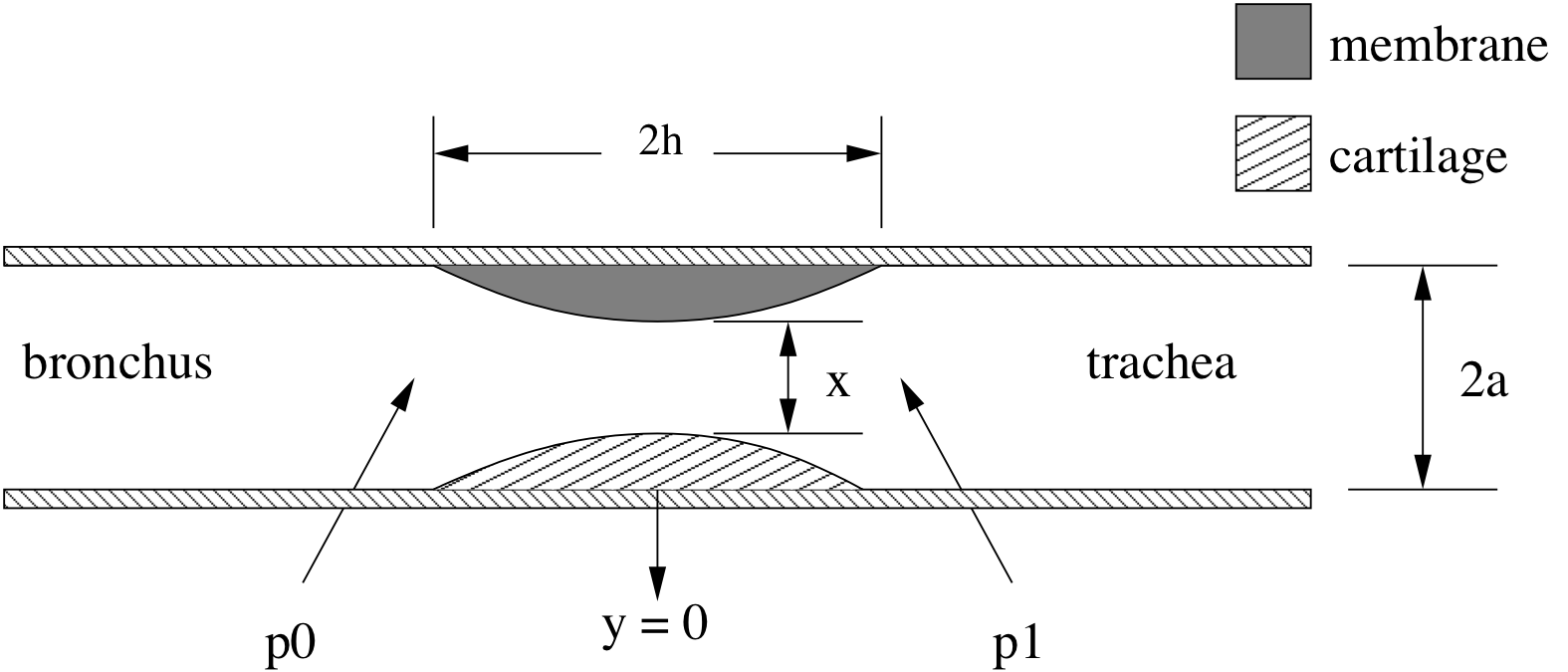, width=3.5in}
\caption{The transverse model of a pressure controlled valve.}
\label{valve}
\end{figure}

\begin{figure*}[ht]
\centering
\epsfig{file=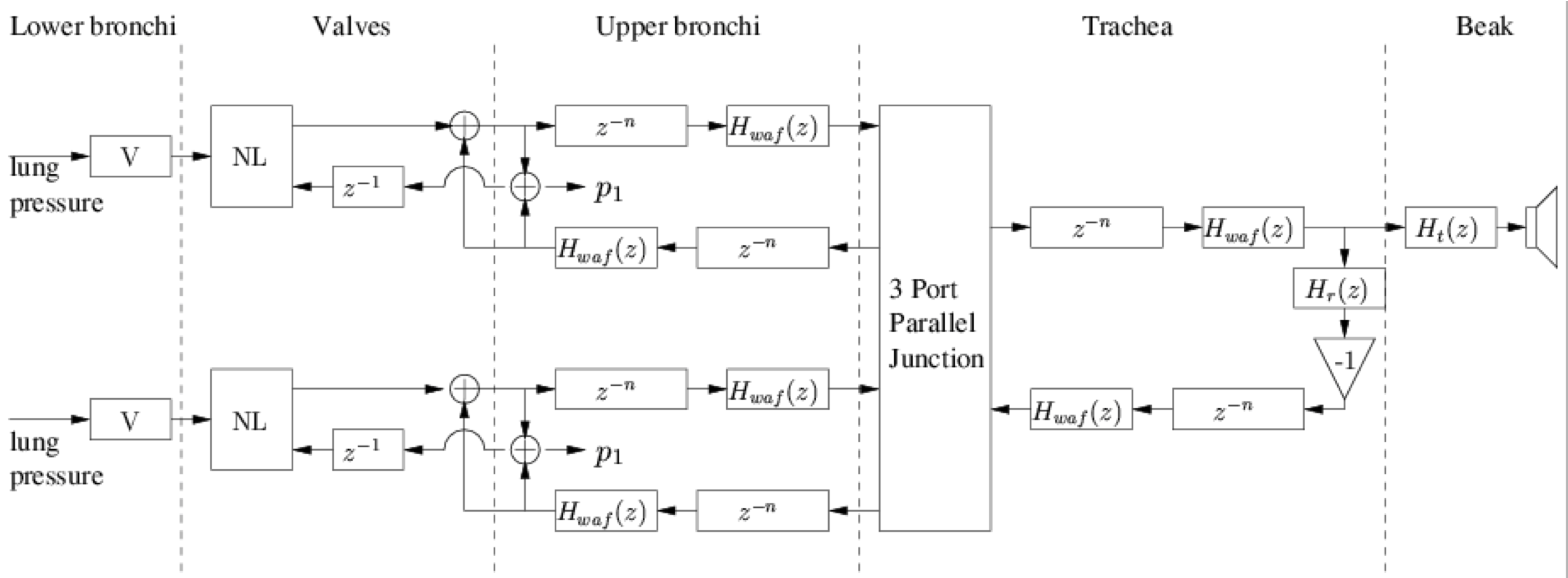, width=6in}
\caption{Signal flow diagram of the model}
\label{model}
\end{figure*}

The valve model has the following four variables which evolve over
time during sound production (illustrated in Fig.~\ref{valve}): the
pressure on the bronchial side of the valve ($p_0(t)$), air volume flow
through the valve channel ($U(t)$), displacement of the membrane ($x(t)$), 
and finally the pressure on the tracheal side of the valve ($p_1(t)$).
The four model variables are simulated by discretizing their
corresponding differential equations, each one very much dependent
on the other.
The methods used for digitally simulating the avian vocal tract model
are described in \cite{smythsmithDAFX02} and 
\cite{smythsmithASA02}.  A signal flow diagram is presented in 
Fig.~\ref{model}.
A physical model of the bird's vocal tract was developed using 
waveguide synthesis techniques for the bronchi and trachea tubes
and finite difference methods for the nonlinear vibrating syringeal
membranes \cite{smythsmithDAFX02,smythsmithASA02}.

The structure of the syrinx varies greatly among different bird
species.  With the anatomical parameters set however, the user is free
to create a tremendous variety of sounds simply by changing the two
primary control parameters: the pressure from the lungs and the
tension in the membrane.

Though there is clearly a definite relationship between the tension in
the membranes and the pitch of the produced sound and likewise between
the blowing pressure and the amplitude, the mapping is made more difficult
by nonlinearities intrinsic to the dynamics of the syrinx \cite{nature98}.
A slight change in one parameter can cause effects such as period
doubling, mode-locking and transitions from periodic to chaotic
behaviour \cite{smythabelsmithSMAC03}.
\section{Mapping}
We are beginning to explore mappings of the mouth shape parameters to the syrinx model parameters,
including left and right syringeal membrane tensions, and 
length and radius of the bronchii and trachea. In addition we plan to try a tactile interface to control
the pressure from the lungs. 
\section{Concluding Remarks}
At present the both the face-tracking mouth controller and a pd implementation of the
syrinx model are up and running. The syrinx model is stable for a range of parameters but
exhibits some instabilities for others.  This is to be expected given
that in reality certain pressure and tension values would only be suited
to certain syrinx geometries and sizes. For instance, blowing pressure 
that is too strong for a small syrinx may cause the membrane or valve to
rupture. Results of our current experiments 
with various mappings will be presented at the conference poster session.
\section{Acknowledgments}
The authors thank Chi-Ho Chan for helpful discussions. The work was supported in
part by the National Institute of Information and Communications Technology of Japan.
%
\bibliographystyle{abbrv}
\bibliography{dslnime04}  

\begin{thebibliography}{10}

\bibitem{brackenbury}
J.~Brackenbury.
\newblock {\em Form and Function in Birds}, chapter Functions of the syrinx and
  control of sound production, pages 193--220.
\newblock New York Academic Press, 1989.

\bibitem{Chan03}
C.~Chan, M.~J. Lyons, and N.~Tetsutani.
\newblock Mouthbrush: Drawing and painting by hand and mouth.
\newblock In {\em Proceedings, ICMI-PUI'03}, pages 277--280, Vancouver, Canada,
  November 2003.

\bibitem{nature98}
M.~S. Fee, B.~Shraiman, B.~Pesaran, and P.~P. Mitra.
\newblock The role of nonlinear dynamics of the syrinx in the vocalization of a
  songbird.
\newblock {\em Nature}, 395:67--71, 1998.

\bibitem{flet88}
N.~H. Fletcher.
\newblock Bird song -- a quantitative acoustic model.
\newblock {\em Journal of Theoretical Biology}, 135:455--481, 1988.

\bibitem{flet99}
N.~H. Fletcher and A.~Tarnopolsky.
\newblock Acoustics of the avian vocal tract.
\newblock {\em Journal of the Acoustical Society of America}, 105(1):35--49,
  January 1999.

\bibitem{Heidegger}
M.~Heidegger.
\newblock {\em The Question Concerning Technology and other Essays}.
\newblock Harper and Row, 1977.

\bibitem{Hewitt03}
D.~Hewitt and I.~Stevenson.
\newblock E-mic: Extended mic-stand interface controller.
\newblock In {\em Proceedings, NIME'03}, pages 122--128, Montreal, Canada, May
  2003.

\bibitem{Lyons03}
M.~J. Lyons, M.~Haehnel, and N.~Tetsutani.
\newblock Designing, playing, and performing with a vision-based mouth
  interface.
\newblock In {\em Proceedings, NIME'03}, pages 116--121, Montreal, Canada, May
  2003.

\bibitem{Lyons01}
M.~J. Lyons and N.~Tetsutani.
\newblock Facing the music: A facial action controlled musical interface.
\newblock In {\em CHI 2001 Extended Abstracts}, pages 309--310, Seattle, USA,
  April 2001.

\bibitem{Merrill03}
D.~Merrill.
\newblock Head-tracking for gestural and continuous control of parameterized
  audio effects.
\newblock In {\em Proceedings, NIME'03}, pages 218--219, Montreal, Canada, May
  2003.

\bibitem{Ng02}
K.~Ng.
\newblock Interactive gesture music performance interfaces.
\newblock In {\em Proceedings, NIME-02}, pages 183--184, Dublin, Ireland, May
  2002.

\bibitem{Orio97}
N.~Orio.
\newblock A gesture interface controlled by the oral cavity.
\newblock In {\em Proceedings, ICMC'97}, pages 141--144, San Francisco, USA,
  1997.

\bibitem{Petajan96}
E.~Petajan and H.~Graf.
\newblock Robust face feature analysis for automation speechreading and
  character animation.
\newblock In {\em Proceedings, FG'96}, pages 357--362, Killington, Vermont,
  USA, 1996.

\bibitem{smythabelsmithSMAC03}
T.~Smyth, J.~Abel, and J.~O. Smith.
\newblock The estimation of birdsong control parameters using maximum
  likelihood and minimum action.
\newblock In {\em Proceedings of SMAC 03}, Stockholm, Sweden, August 2003.

\bibitem{Smyth02}
T.~Smyth and J.~O. Smith.
\newblock Creating sustained tones with the cicada's rapid sequential buckling
  mechanism.
\newblock In {\em Proceedings, NIME'02}, Dublin, Ireland, May 2002.

\bibitem{smythsmithDAFX02}
T.~Smyth and J.~O. Smith.
\newblock The sounds of the avian syrinx---are they really flute-like?
\newblock In {\em Proceedings, DAFX'02}, Hamburg, Germany, September 2002.

\bibitem{smythsmithASA02}
T.~Smyth and J.~O. Smith.
\newblock The syrinx: Nature's hybrid wind instrument.
\newblock In {\em CD-ROM Paper Collection}, Cancun, Mexico, September 2002.

\bibitem{Vogt02}
F.~Vogt, G.~McCaig, M.~A. Ali, and S.~Fels.
\newblock Tongue 'n' groove: An ultrasound based music controller.
\newblock In {\em Proceedings, NIME'02}, pages 60--64, Dublin, Ireland, May
  2002.

\end{thebibliography}
\balancecolumns
%
%

\end{document}